# Reducing parasitic capacitance of strained Si nano p-MOSFET by control of virtual substrate doping

Mohammad Mahdi Khatami[a], Majid Shalchian[a]*, Mohammadreza Kolahdouz[b]

[a]Department of Electrical Engineering, Amirkabir University of Technology, Tehran, Iran
[b]Department of Electrical and Computer Engineering, University of Tehran, Tehran, Iran
*shalchian@aut.ac.ir

**Abstract:** Biaxially strained Si channel p-MOSFETs on virtual SiGe substrates suffer from a parasitic parallel channel in virtual substrate. This channel participates in current passing through the devices and increases parasitic capacitors which degrades the high frequency response. In this paper a new approach has been introduced to eliminate the channel which in turn reduces parasitic capacitors of the MOSFET. It is illustrated that, increasing virtual substrate doping, can reduce and finally eliminate this unintentional channel. In this work 2D simulation has been used to investigate the impact of the proposed method.

**Keywords:** biaxially strained Si p-MOSFET, relaxed, virtual substrate, quantum well, parasitic capacitors

## Introduction

As semiconductor industry is following Moore's law, it faces some technology red brick walls that make trouble in its way, e.g. are thermal dissipation, leakage current, and so on [1], [2]. Thus new approaches are required to solve these issues.

It has been shown that biaxially strained Si can improve the carrier's mobility[3], [4]. Therefore using strained Si as the MOSFET's channel, reduces the channel resistance. However, there is a major drawback in this approach for p-MOSFETs. Difference in the band energy levels between strained Si and SiGe layer forms a quantum well (QW) for holes within SiGe layer. This QW confines holes inside SiGe layer instead of strained Si. This phenomenon increases leakage current and parasitic capacitors, which degrades MOSFET's ideality.

One solution proposed by Rim *et al* [5] and analyzed by Maiti and Armstrong [6], was to use a graded SiGe layer between strained Si and relaxed SiGe. Ge fraction of the graded layer changes from zero to Ge fraction of relaxed SiGe. Although this method removes parasitic channel, it simultaneously makes strained layer thicker which results in more defects in the layer, so it is not desired.

Other solution proposed by Sugii *et al* [7], was based on increasing the doping of a part of the strained Si channel to eliminate the parasitic channel. But this method makes fabrication process complicated and also increases the number of defects.

In this work a novel approach has been proposed using doping engineering in virtual substrate of the p-MOSFETs to passivate the parasitic channel. 2D Poisson numerical simulation is used to prove the impact of this method on eliminating the parasitic channel and to reduce the parasitic capacitance. This work is based on the more recent models of the strained Si and SiGe which proposed in [8]–[12].

## Structure

A cross section view of the MOSFET has been shown in Figure 1. The structure is based on growing a thin Si (15 nm) on a thick fully relaxed SiGe. As lattice constant of SiGe is larger than Si (almost 4.2%), the Si will be strained in two dimensions. To have fully relaxed SiGe with good quality, it should be a graded SiGe layer between SiGe layer and Si substrate[13]. The channel length is assumed to be 100 nm with doping concentration of $4 \times 10^{17}\,\text{cm}^{-3}$. A p-poly is used as the MOSFET's gate, and the oxide thickness is set to 4 nm. Si lattice orientation is (100).

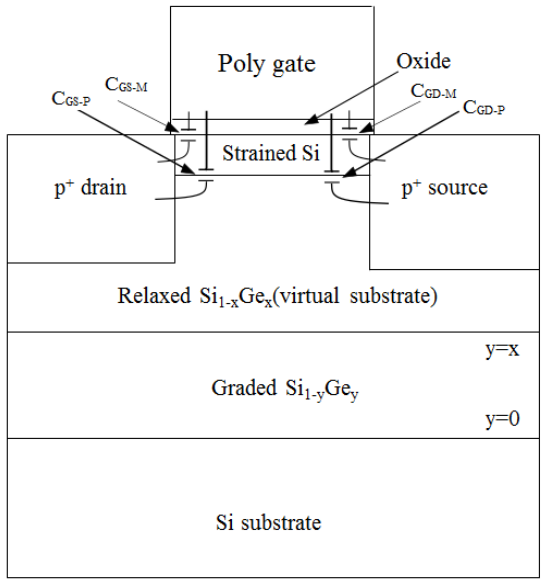

Figure 1 cross section view of a biaxially strained p-MOSFET

## Results and Discussion

There are two major parasitic capacitors which affect high frequency response of MOSFET: gate-source capacitor and gate-drain capacitor. The structure under review has been studied for three doping levels of virtual substrate: $4 \times 10^{15}$ cm$^{-3}$ as the first case, $4 \times 10^{16}$ cm$^{-3}$ as the second case, and $4 \times 10^{17}$ cm$^{-3}$ as the third case.

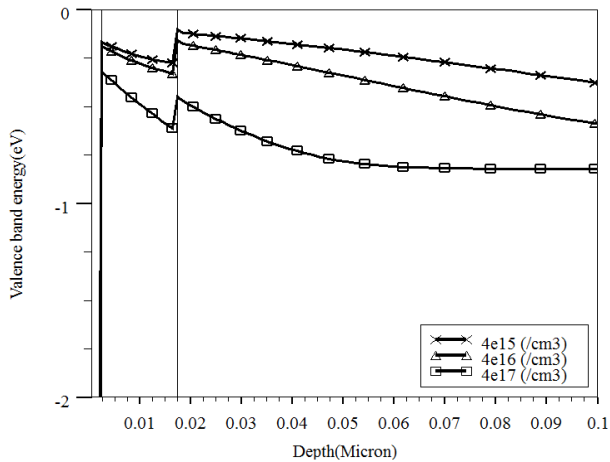

Figure 2 Valence bands alignment for the three cases of SiGe doping

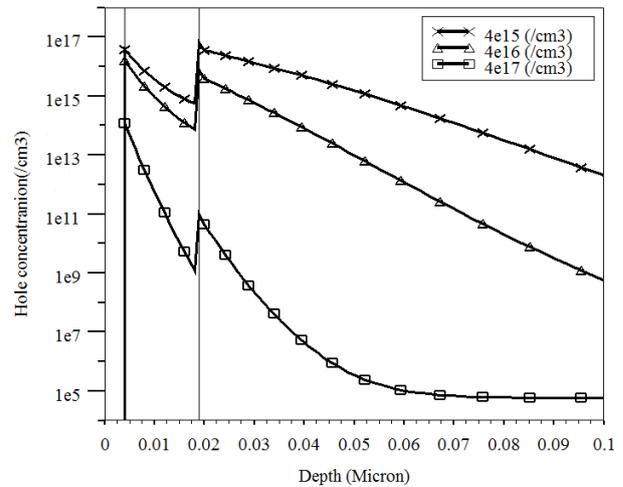

Figure 3 Hole concentration for the three cases of SiGe doping

Figure 2 illustrates the valence band of the MOSFET's layers in zero bias ($V_{SG}$=0 V, $V_{SD}$=0). In light SiGe doping, valence band level of SiGe is higher than strained Si, which causes holes to occupy SiGe energy states. This means that the SiGe acts as a quantum well for holes. This leads to formation of a parasitic channel in SiGe parallel to the strained Si channel which contribute in passing current. Increasing SiGe doping reduces the width of the quantum well and decreases the number of the energy levels and the barrier height of QW.

So increasing the SiGe's doping level in this device, diminishes the role of the quantum well. If doping is large enough, then the current will flow just through the strained Si.

Figure 3 also illustrates hole concentration of the structure in zero bias condition. As illustrated in the third case that doping concentration is $4 \times 10^{17}$ cm$^{-3}$, the density of excess hole in the parasitic channel has been considerably reduced.

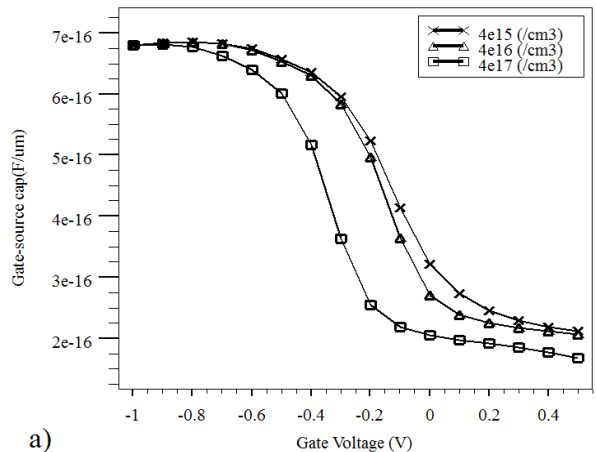

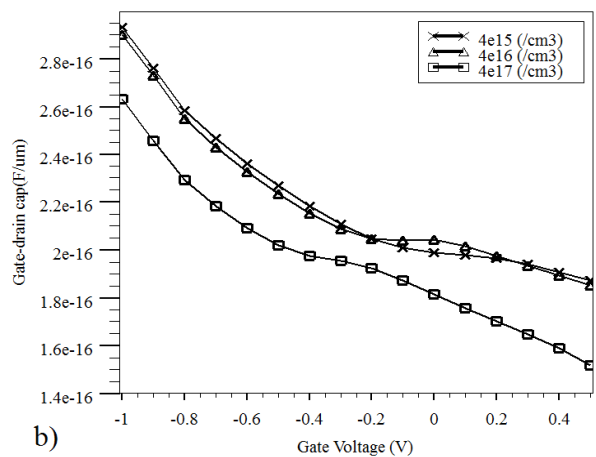

**Figure 4** Parasitic capacitors a function of $V_{GS}$ for the three cases of SiGe doping. a) gate-source capacitor b) gate-drain capacitor

Figure 4-a and 4-b illustrate gate-source and gate-drain parasitic capacitances ($C_{GS}$ and $C_{GD}$) as a function of $V_{SG}$ ($V_D=0$ and $V_{SD}=0.4$ V). A first order model of internal capacitors within the device has been demonstrated in Figure 1. It is expected that gate to source capacitor has two parallel components, one is controlled by main channel in silicon specified by $C_{GS-M}$ and other one is controlled by parasitic channel in SiGe specified by $C_{GS-P}$, therefore it is expected that by reducing parasitic channel, the related component of gate-source capacitance also would be reduced, which results in lower total capacitance from gate to source. With similar thought $C_{GD}$ would be reduced with eliminating parasitic channel. This explain the characteristics of Figure 4-a and Figure 4-b.

**Conclusion**

In this paper we have demonstrated that by controlling the doping of virtual channel, formation and contribution of parasitic channel on MOSFETs can be controlled. More specifically for doping level of $4 \times 10^{17}$ cm$^{-3}$ parasitic channel could be eliminated which results in lower parasitic capacitances.

**Acknowledgment**

The authors would like to thank late Dr. Saeid Khatami, for his great help and fruitful discussions.